\documentclass[11pt,a4paper]{article}
\usepackage{amsmath,amsthm,amssymb}
\usepackage{array}
\usepackage{geometry}
\usepackage{hyperref} 
\usepackage{epsfig}
\usepackage{graphicx,color}
\usepackage{float}
\usepackage{multirow}
\theoremstyle{plain}

\theoremstyle{definition}

\numberwithin{equation}{section}

\begin{document}

\title{EPI-KAN: A Method For Estimating and Forecasting Time-Dependent COVID-19 Parameters}

\author{Arief Anbiya \\ Independent Researcher, Indonesia \\ \texttt{ariefanbiya@gmail.com}}
\date{\today}
\maketitle

\begin{abstract}
We introduce EPI-KAN, a novel method for estimating COVID-19 time-varying parameters. EPI-KAN uses historical epidemiological data, Physics-Informed Neural Network (PINN), and the novel Kolmogorov-Arnold Network (KAN). The method harnesses the novel Kolmogorov-Arnold Network (KAN), which is a type of artificial neural network. For the KAN in this paper, we learn activation functions that are represented using Fourier series, hence we abbreviate as KAN-F. In this study, we estimate parameters in the context of an SIRD compartmental differential equations. The time-dependent parameters are the transmission rate $\beta(t)$, recovery rate $\gamma(t)$, and mortality rate $\mu(t)$. We define three KAN-F functions $\widehat{\beta}$, $\widehat{\gamma}$, $\widehat{\mu}$ that model the true parameters $\beta(t)$, $\gamma(t)$, $\mu(t)$, respectively. We test two model architectures for the KAN-F: the first has 8 input variables consisting of $S$, $I$, $R$, $D$, and their numerical gradients at any time $t$, while the second has 4 input variables excluding the numerical gradients. The objective loss function that has to be minimized is subject to Physics-Informed Neural Network (PINN). Using historical data of COVID-19 from three South-East Asian countries: Indonesia, Singapore, and Malaysia, we are able to estimate $\beta(t)$, $\gamma(t)$, and $\mu(t)$ on each country with decent accuracy and efficiency. The time period of choice coincides with the period where SARS-CoV-2 Delta variant (B.1.617.2) was dominant. In addition to estimating the rates during the training period, we also predict transmission rates over 30 days during forecast period. We found that the output of KAN-F over the forecast period can give good predictions if we scale the output by a factor of 17\% for Indonesia and 30\% for Singapore and Malaysia.

\end{abstract}






\section{Introduction}\label{sec:intro}
\noindent 
The modeling and analysis of infectious disease dynamics plays an important role in informing public health responses, designing intervention strategies, and anticipating healthcare service demands. Compartmental models, in particular, are useful tools in this effort. Compartmental models divide the population into different categories and describe the transfer of people between categories using systems of ordinary differential equations or even partial differential equations. By calibrating compartmental models with observed data, one can simulate the trajectory of an outbreak and evaluate the potential effects of interventions. However, the accuracy of compartmental models is highly dependent on the relevance of their parameters. Traditional compartmental models often rely on constant parameters, assuming that transition rates remain fixed throughout the course of an outbreak. This assumption can fail to capture the complexities of real-world epidemics. For instance, transmission rates can be influenced by government interventions, behavioral changes in the population, emergence of new variants, and seasonality. Similarly, recovery rates can be influenced by evolving treatment protocols, healthcare availability, and increased vaccine coverage.

To better reflect the intricate nature of actual epidemic scenarios, time-dependent parameters can be applied in epidemic modeling. By allowing parameters to evolve as functions of time, models can more accurately reproduce the nonlinear behavior observed in disease outbreaks. The challenge lies in determining the appropriate time-dependent parameters. One way to construct parameter functions is to find those such that the model’s solution aligns with real-world data. This can be considered as an inverse problem: we can treat the real-world data as a solution to a system of differential equations (epidemic model), and then attempt to identify the time-dependent parameters such that the differential equations have solutions tantamount to the real-world data. 

In this study, we address this challenge by harnessing historical epidemiological data of COVID-19, Physics-Informed Neural Network (PINN)  \cite{pinn_many_cites}, and the novel Kolmogorov-Arnold Network (KAN) \cite{kolmogorov_arnold_network} to estimate time-dependent COVID-19 parameters. We adopt the abbreviation EPI-KAN. KAN is a type of artificial neural network that has shown promising capabilities in applications such as the fitting of data and the solving of partial differential equations, when compared to the more prevalent Multi Layer Perceptron (MLP) \cite{kolmogorov_arnold_network}. The time-dependent COVID-19 parameters will be attached to a Susceptible–Infective–Recovered-Deceased (SIRD) compartmental model (\ref{sird_model_S})-(\ref{sird_model_D}). The SIRD model involves three time-dependent epidemic parameters: the transmission rate $\beta(t)$, recovery rate $\gamma(t)$, and mortality rate $\mu(t)$. Throughout the paper, we may use the term ``rates'' instead of ``parameters'', whenever we refer to $\beta(t)$, $\gamma(t)$, and $\mu(t)$. We will use KAN functions, in combination with Fourier series, to model (estimate) these time-dependent COVID-19 parameters. To avoid confusion between time-dependent parameters of compartmental differential equations and trainable parameters of a neural network model, we adopt the term NN-parameters to refer to neural network trainable parameters. In traditional MLP, we have fixed activation functions and must learn a set of weight-coefficients and bias-constants as the learnable NN-parameters. In KAN, we have single-variable activation functions that must be learned, and these are the only trainable objects. Although the activation functions learned in KAN are commonly approximated using spline functions, we will use Fourier series to represent the activation functions. Hence, in this paper, we introduce the abbreviation KAN-F to refer to KAN that uses Fourier series as proxy to the learnable activation functions. Each Fourier series has a number of learnable wave amplitudes and frequencies. The NN parameters in KAN-F is the set of all learnable Fourier amplitudes and frequencies of all the activation functions (Fourier series). 

We will estimate the true rate functions $\beta(t), \gamma(t), \mu(t)$ using three different KAN-F functions with specific architectures. These network functions will be denoted as $\widehat{f_{\beta}}(\mathbf{x}[t])$, $\widehat{f_{\gamma}}(\mathbf{x}[t])$, and $\widehat{f_{\mu}}(\mathbf{x}[t])$, which are subsequently be mapped to the sigmoid function $Sigmoid(x)=\frac{e^{x}}{1+e^{x}}$ to obtain $\widehat{\beta}(\mathbf{x}[t])=Sigmoid(\widehat{f}_{\beta}(\mathbf{x}[t]))$, $\widehat{\gamma}(\mathbf{x}[t])=Sigmoid(\widehat{f}_{\gamma}(\mathbf{x}[t]))$, and $\widehat{\mu}(\mathbf{x}[t])=Sigmoid(\widehat{f}_{\mu}(\mathbf{x}[t]))$. The functions $\widehat{\beta}$, $\widehat{\gamma}$, $\widehat{\mu}$ will be the estimate for the true COVID-19 parameters $\beta(t), \gamma(t), \mu(t)$. The definition of $\mathbf{x}[t]$ is explained in more detail in Section \ref{sec:kan_fourier_specific}. We develop two different model architectures and define them as Model A and Model B. For Model A, the input vector $\mathbf{x}[t]$ will consist of 8 input variables: the 4 state variables $S$, $I$, $R$, $D$ and their rate of change at some time point $t$. For Model B, the input vector $\mathbf{x}[t]$ will consist of only 4 input variables: the 4 state variables $S$, $I$, $R$, $D$ without including their rate of change at some time point $t$. Although the approach of using a temporal variable $t$ as a single input is predominant in the context of PINN \cite{cheng_dinn,xiao_fourier_mapping,pinn_sir,pinn_seir_estimate_E,pinn_hospitalization,fourier_mapping_1,Nguyen_Raisi_PINN,Kharazmi_PINN, Ning_PINN_Euler, Han_PINN, Shamsara_PINN,Jie_PINN, NING_PINN_MDPI,He_PINN,oluwasakin_PINN_2023,olumoyin_PINN_2021}, we found that our method of using multiple relevant-inputs is comparatively more efficient without dispensing accuracy. Intuitively, an ``artificial brain'', which refers to a neural network model, would learn to estimate epidemic parameters better if it were given meaningful information that are relevant to the disease dynamics, rather than merely a single input of time variable $t$.

In \cite{xiao_fourier_mapping}, the same SIRD compartmental model and its parameters were also used for time-dependent parameters estimation. In \cite{xiao_fourier_mapping}, they used neural networks of the type MLP, combined with Fourier feature-mapping in the input layer, and Residual Network (ResNet) \cite{res_net_paper} to model the time-dependent epidemic parameters. In contrast, our study uses the novel KAN-type neural network with Fourier series representation. Note that the Fourier feature-mapping in \cite{xiao_fourier_mapping} and the Fourier series representation in our paper are entirely different concepts and methods. Also inspired by \cite{xiao_fourier_mapping}, we slightly modify their objective loss functions that have to be minimized, which are the mean-of-squared errors (MSE) between the exact solutions of the SIRD model against its Runge-Kutta \cite{runge_kutta} numerical solutions at discrete time points. Instead of using MSE, we use root-mean-of-squared-errors (RMSE) (\ref{loss_S})-(\ref{loss_D}). The exact solutions of the SIRD model (at discrete time points) are taken as real COVID-19 data.  This objective loss function must be minimized with respect to many NN-parameters (Fourier amplitudes and frequencies) of the KAN-F functions $\widehat{\beta}$, $\widehat{\gamma}$, and $\widehat{\mu}$. 

This approach is basically the same as Physics Informed Neural Network (PINN) \cite{pinn_many_cites, bayesian_pinn}, but the differential equations in our case are based on epidemiological compartment model that categorizes more to social and biological dynamics instead of natural law of physics. This approach combines the capability of deep neural networks with the constraints imposed by differential equations of infectious-disease model. Although much of the literature also uses the term PINN for epidemic model parameter identification \cite{pinn_sir,pinn_seir_estimate_E,pinn_hospitalization,fourier_mapping_1,Nguyen_Raisi_PINN,Kharazmi_PINN, Ning_PINN_Euler, Han_PINN, Shamsara_PINN,Jie_PINN, NING_PINN_MDPI,He_PINN,pinn_africa}, this paper adopts the term Epi-DNN instead of PINN, since the context is of epidemic-related differential equations. The term Epi-DNN was also adopted in \cite{xiao_fourier_mapping}. 

\begin{figure}
    \centering 
\makebox[\textwidth]{
\includegraphics[width=18cm]{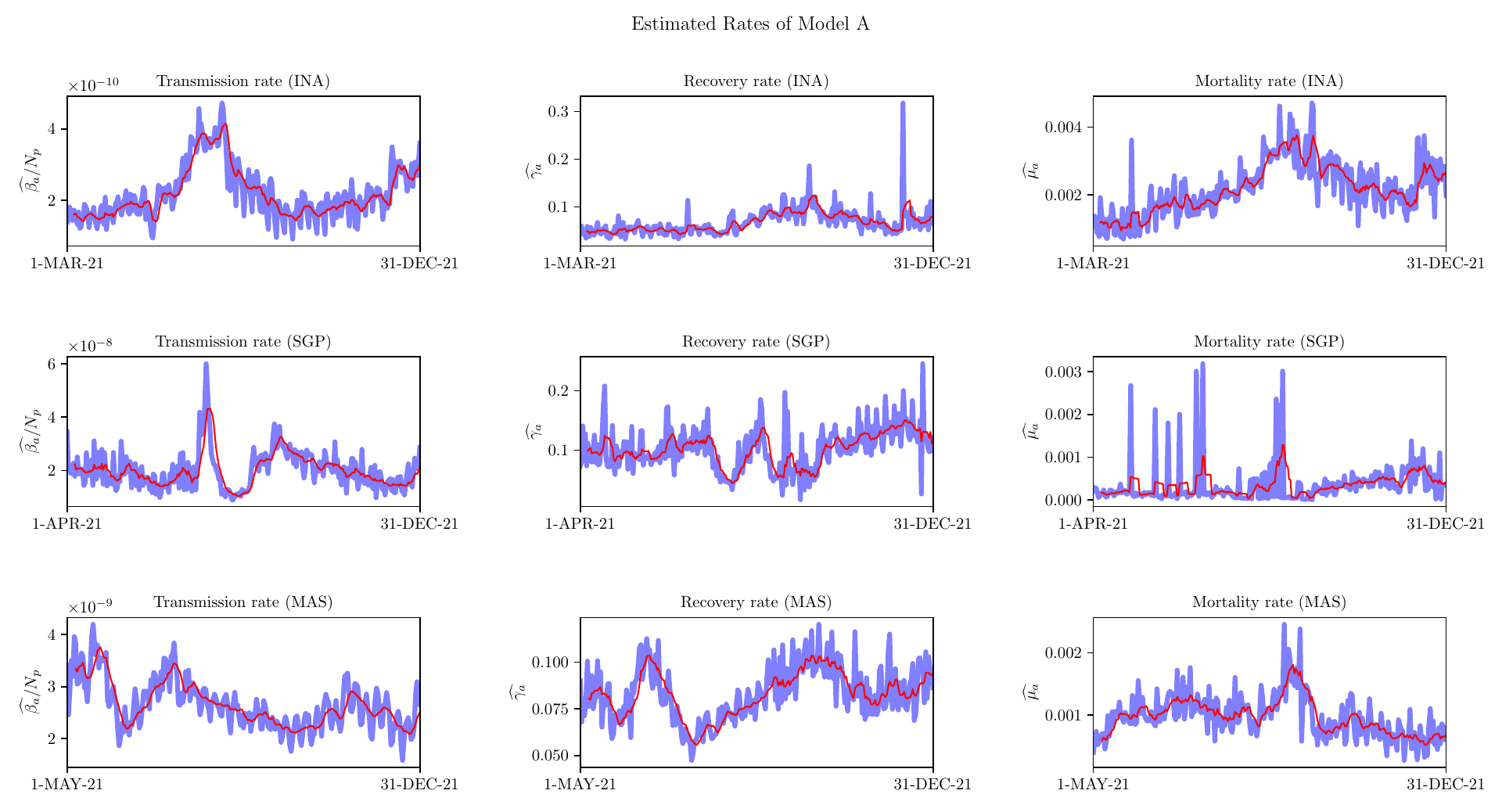}}
    \caption{The estimated rates (blue) for Indonesia (INA), Singapore (SGP), and Malaysia (MAS). The red lines are the seven-day moving averages (7-MA).}
\label{fig:estimated_rates_delta}
\end{figure}

\begin{figure}
    \centering 
\makebox[\textwidth]{
\includegraphics[width=15cm]{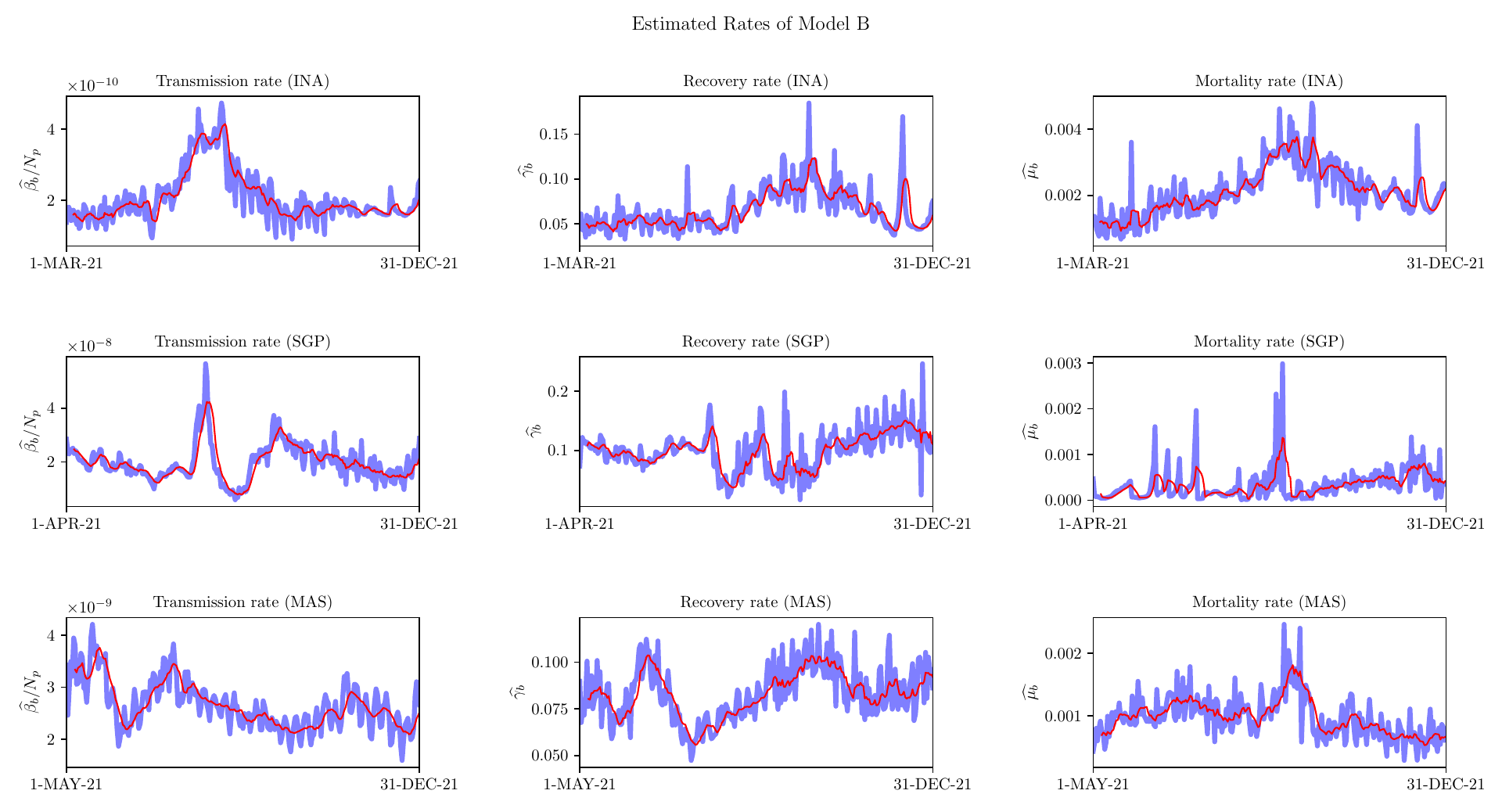}}
    \caption{The estimated rates (blue) according to Model B for Indonesia (INA), Singapore (SGP), and Malaysia (MAS). The red lines are the seven-day moving averages (7-MA).}
\label{fig:estimated_rates_delta_PREDICTOR}
\end{figure}

The aim of this paper is to usher in a KAN-F application for COVID-19 time-dependent parameters estimation. We used the same compartmental model, parameters, and virtually the same objective loss functions as in \cite{xiao_fourier_mapping}, in order to present relevant comparison. We found that our KAN-F models can estimate the time-dependent parameters accurately and more efficiently: the data-fittings are almost exact under a longer COVID-19 time period and with significantly fewer number of training steps (epochs). The final estimated rates are plotted in Figure \ref{fig:estimated_rates_delta} and Figure \ref{fig:estimated_rates_delta_PREDICTOR}. The estimated transmission, recovery, and mortality rates that are modelled according to Model A are written as $\widehat{\beta_{a}}$, $\widehat{\gamma_{a}}$, and $\widehat{\mu_{a}}$, respectively. Similarly for Model B, they are written as $\widehat{\beta_{b}}$, $\widehat{\gamma_{b}}$, and $\widehat{\mu_{b}}$. Throughout this paper, the general notations that can either refer to Model A or Model B are $\widehat{\beta} \in \{ \widehat{\beta_{a}}, \widehat{\beta_{b}}\}$, $\widehat{\gamma} \in \{ \widehat{\gamma_{a}}, \widehat{\gamma_{b}}\}$, and $\widehat{\mu} \in \{ \widehat{\mu_{a}}, \widehat{\mu_{b}}\}$. The model-fitting resulted from these estimated rates are shown in Figure \ref{fig:model_fitting_all} and Figure \ref{fig:model_fitting_all_PREDICTOR}, while the fitting errors are summarized in Table \ref{table:model_fitting_all} and \ref{table:model_fitting_all_PREDICTOR}. The training evolutions (optimization of the loss functions) are shown in Figure \ref{fig:log_loss}. We are able to obtain good estimate of the rates with less than $10^{4}$ training epochs. This is significantly more efficient than the result in \cite{xiao_fourier_mapping}, which requires $10^{5}$ epochs to estimate COVID-19 parameters in a shorter time period of three months using China COVID-19 dataset. Here, we estimate COVID-19 parameters for three countries in South-East Asia: Indonesia, Singapore, and Malaysia, over a period of at least seven months. 

The novel combination of KAN and Fourier series has been discussed recently in \cite{kolmogorov_arnold_fourier_network}, but not in the context of infectious disease modelling. To the best of our knowledge, there has been no published article discussing the applications of KAN and Fourier series in COVID-19 parameters estimation. Although there is a recent study that uses KAN to fit a classical SIR infectious disease model \cite{kan_covid_1}, they used constant SIR model parameters and a simulated dataset. However, we are aware of similar studies on COVID-19 parameters estimation. In \cite{xiao_fourier_mapping}, they used an SIRD model with three epidemic parameters, all are time-dependent. In \cite{Shamsara_PINN}, they used an SIHRD model with seven time-dependent epidemic rates, but the rates are estimated using MLPs as constants for different time windows (each time window requires separate neural network training). They subsequently took the average of these constants from different and intersecting time windows to obtain the rates, for each of the time points. Many other studies involving Epi-DNN also estimated epidemic parameters \cite{pinn_sir,pinn_seir_estimate_E,pinn_hospitalization,fourier_mapping_1,Nguyen_Raisi_PINN,Kharazmi_PINN, Ning_PINN_Euler, Han_PINN,Jie_PINN, NING_PINN_MDPI,He_PINN,oluwasakin_PINN_2023,olumoyin_PINN_2021}. Studies excluding neural networks for COVID-19 parameters estimation are \cite{levenberg_marquardt,CBMS_WIJAYA,CBMS_NADIA,HONG_notNN,IMA_Luo,IOP_Sikder,BATISTA_2022,CARMEN_MDPI,adaptive_SIR, MVI_1, MVI_2, MVI_3, periodic_lockdown_asirv,periodic_lockdown_asirv_2}. However, none of those studies involve regular KAN or even KAN-F. Moreover, our paper uses up to 8 input variables for the networks, which involves 4 state variables $S,I,R,D$ and their numerical gradients. This is relatively new, compared to the existing literature that uses time $t$ as a single input \cite{cheng_dinn,xiao_fourier_mapping,pinn_sir,pinn_seir_estimate_E,pinn_hospitalization,fourier_mapping_1,Nguyen_Raisi_PINN,Kharazmi_PINN, Ning_PINN_Euler,Han_PINN, Shamsara_PINN,Jie_PINN, NING_PINN_MDPI,He_PINN,oluwasakin_PINN_2023,olumoyin_PINN_2021}.

The content in this paper is structured as follows:
\begin{itemize}
\item In Section \ref{sec:kan_intro}, we briefly introduce and discuss the Kolmogorov-Arnold representation theorem, Kolmogorov-Arnold Network, and Fourier series.
\item In Section \ref{sec:sird_definition}, we describe the SIRD compartmental differential equations and their time-dependent parameters.
\item In Section \ref{sec:data_preparation}, we explain how to obtain COVID-19 data for $S$, $I$, $R$, and $D$ variables. In Section \ref{sec:data_preparation_NN}, we explain how to construct the neural network training data based on the COVID-19 data.
\item In Section \ref{sec:kan_fourier_specific}, we briefly describe the KAN-F architecture that we apply and found to have decent performance in estimating the time-dependent rates. 
\item In Section \ref{sec:loss_function}, we explain the objective loss function that has to be minimized with respect to the NN parameters.
\item In Section \ref{sec:results}, we summarize the neural network training results for each of the three countries. 
\item In Section \ref{sec:forecast},  in addition to estimating the rates over the training period, we also predict transmission rates over 30 days outside of training period.
\end{itemize}

\section{Kolmogorov-Arnold Network with Fourier series} \label{sec:kan_intro}

\subsection{Kolmogorov-Arnold Representation Theorem}

\noindent Let $\mathcal{I}=[0,1] \subset \mathbb{R}$ be the unit interval. Kolmogorov-Arnold Representation Theorem \cite{kolmogorov_arnold_theorem_1} states that any multivariate continuous function $f: \mathcal{I}^{n} \rightarrow \mathbb{R}$ can be written as the following representation:
\begin{equation}  \label{kolmogorov_arnold_function_0}
f(x_{1},x_{2},\hdots,x_{n}) = \sum_{i=1}^{2n + 1} \Phi_{i} \left( \sum_{j=1}^{n} \phi_{i,j} \left( x_{j} \right) \right). 
\end{equation}
The functions $\phi_{i,j}$s are called the inner functions and each of them is continuously defined on $\mathcal{I}$, and the functions $\Phi_{i}$s are called the outer functions and they are continuous. A proof of this theorem can be seen in \cite{kolmogorov_arnold_theorem_2}.

We can extend this representation to any arbitrary interval $\mathcal{C}=[a,b] \subset \mathbb{R}$. Any variable $x_{i}$ defined on $\mathcal{C}$ can be written as $x_{i} = a + y_{i}(b-a)$, where $y_{i}$ is another variable defined on $\mathcal{I}$.
Hence, a continuous function $g: \mathcal{C}^{n} \rightarrow \mathbb{R}$ can be written as
\begin{align*}  
g(x_{1},x_{2},\hdots,x_{n}) &= g(a + y_{1}(b-a), a + y_{2}(b-a), \hdots, a + y_{n}(b-a))  \\
&= f(y_{1}, y_{2}, \hdots, y_{n}) \\
&= \sum_{i=1}^{2n + 1} \Phi_{i} \left( \sum_{j=1}^{n} \phi_{i,j} \left( y_{j} \right) \right),
\end{align*}
so that we have
\begin{equation}\label{kolmogorov_arnold_function} 
g(x_{1},x_{2},\hdots,x_{n}) =  \sum_{i=1}^{2n + 1} \Phi_{i} \left( \sum_{j=1}^{n} \phi_{i,j} \left( (x_{j} - a)/(b-a) \right) \right). \end{equation}

\begin{figure}
    \centering
\makebox[\textwidth]{
\includegraphics[width=10cm]{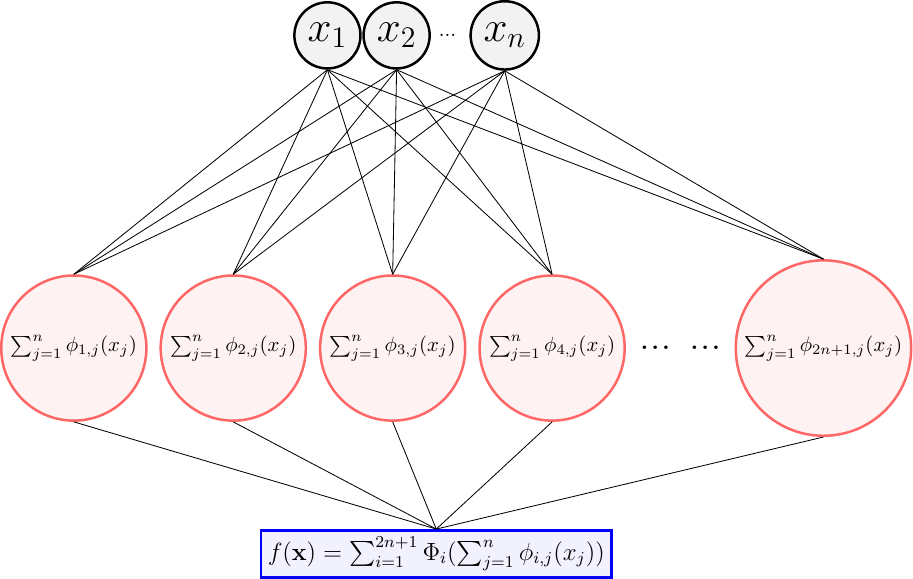}}
    \caption{Network illustration of Kolmogorov-Arnold Representation Theorem with layer-sequence $[n,2n+1,1]$.}
\label{fig:original_KAN_architecture}
\end{figure}

\subsection{Kolmogorov-Arnold Network}
\noindent The Kolmogorov-Arnold function representation (\ref{kolmogorov_arnold_function_0}) or (\ref{kolmogorov_arnold_function}) can be viewed as a neural network model with an input layer of $n$ neurons, a hidden layer with $2n+1$ neurons, and an output layer with 1 output neuron. Figure \ref{fig:original_KAN_architecture} illustrates an aritificial neural network with 1 hidden layer based on Kolmogorov-Arnold representation. Between the input layer and the hidden layer, we will learn inner functions $\{\phi_{i,j} | i = 1,\hdots,2n+1, j = 1,\hdots, n\}$, and then use them so that the sum $\sum_{j=1}^{n} \phi_{i,j}(x_{j})$ will be the value to be used in the $i_{\text{th}}$ neuron in the hidden layer. The final step is the transfer from the hidden layer to the output layer (of single output neuron) by learning the outer functions $\{\Phi_{i} | i=1,\hdots,2n+1 \}$ and have the addition $\sum_{i=1}^{2n+1} \Phi_{i}(\sum_{j=1}^{n} \phi_{i,j}(x_{j}))$ as the value to be used in the final output neuron. We can say that the representation \ref{kolmogorov_arnold_function_0}) or (\ref{kolmogorov_arnold_function}) is a network with layer-sequence $[n,2n+1,1]$ of the number of neurons in the input layer, the hidden layer, and the output layer, respectively. Unlike the traditional MLP, where we train the network by finding the best weight-coefficients and bias-constants, in KAN we train the network by finding the best inner and outer functions that minimizes an objective loss function. We may generalize KAN to deeper networks \cite{kolmogorov_arnold_network}, see Figure \ref{fig:KAN_architecture}.

We will give an example where there is more than one hidden layer. For example, we may have a model with layer-sequence $[5,11,11,1]$. Let the input variables be stored as $\mathbf{x}=[x_{1},\hdots,x_{n}]$. The Kolmogorov-Arnold Network model evaluated at $\mathbf{x}$ is denoted as $\widehat{f}(\mathbf{x})$, and computed as follows:

\begin{itemize}
\item The transfer from the input layer to the first hidden layer:
$$ Z_{j}^{(1)} =  \sum_{k=1}^{5} \phi_{j,k}^{(1)}(x_{k}) ,   $$
for $j=1,\hdots,11$.
\item  The transfer from the first hidden layer to the second hidden layer:
$$ Z_{j}^{(2)} = \sum_{k=1}^{11} \phi_{j,k}^{(2)}(Z_{k}^{(1)}) ,   $$
for $j=1,\hdots,11$.
\item The transfer from the second hidden layer to the output layer with single neuron: $$ \widehat{f}\left(\mathbf{x}\right) = \sum_{k=1}^{11} \phi_{1,k}^{(3)}(Z_{k}^{(2)}) $$
\end{itemize}

In general, if we have $n$ input variables, with $M$ hidden layers, with each hidden layer to have $N(i)$ neurons, for $i=1,\hdots,M$, and only one output neuron, then the KAN model with these hyperparameters can be depicted as in Figure \ref{fig:KAN_architecture}.

\subsection{Deep Kolmogorov-Arnold Network Supported by Representation Theorem}

\begin{figure}
    \centering
\makebox[\textwidth]{
\includegraphics[width=10cm]{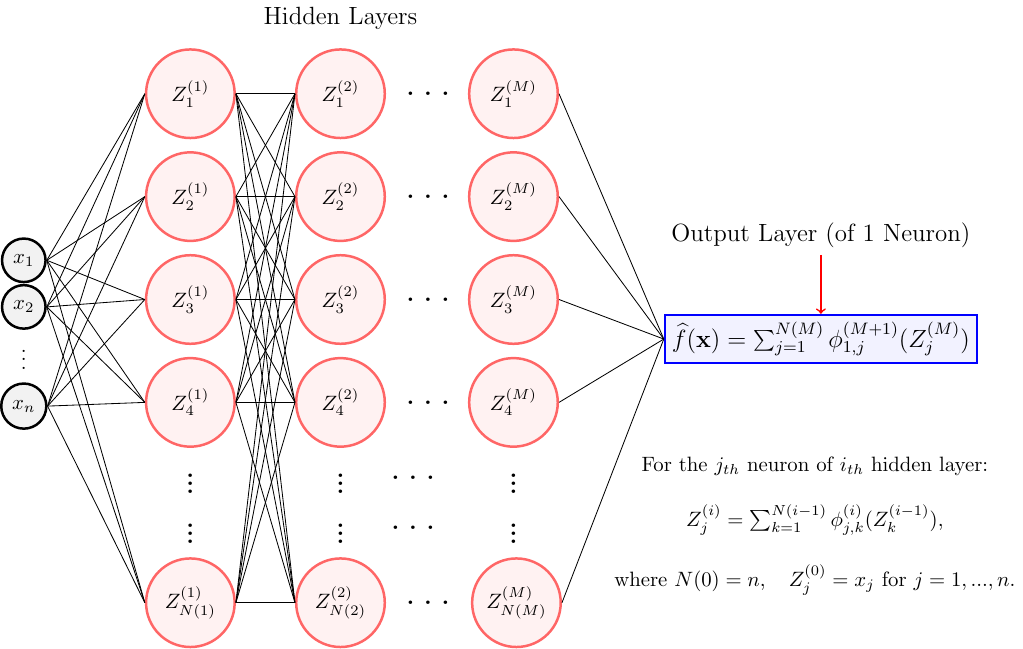}}
    \caption{Illustration of a KAN with architecture-sequence $[n,N(1),\hdots,N(M),1]$.}
\label{fig:KAN_architecture}
\end{figure}

\noindent Based on the Kolmogorov-Arnold representation theorem \cite{kolmogorov_arnold_theorem_1,kolmogorov_arnold_theorem_2}, only a network with 1 hidden layer that is guaranteed to be a ``universal'' function approximator. Moreover, if there are $n$ input features, the number of neurons of the single hidden layer must be $2n + 1$. Thus, if we want to use numerous hidden layers, we can treat the final two hidden layers as an implicit input layer and an implicit single hidden layer, respectively, of a basic KAN. Using Figure \ref{fig:KAN_architecture} as reference, the $(M-1)_{th}$ and $M_{th}$ layers can be seen as an implicit input layer and an implicit single hidden layer, respectively, of a basic KAN. We require the number of neurons in the $M_{th}$ layer to be $N(M) = 2N(M-1)+1$, so that the last two hidden layers and the output neuron are consistent with the theorem. We can view the true input layer, together with the 1st up to $(M-2)_{th}$ hidden layers, as a network that computes for the best ``input'' values in the basic KAN that starts at $(M-1)_{th}$ layer.

\subsection{Fourier Series Approximation}
\noindent Any continuous single variable function $f(x)$, defined on an interval $[-L, L] \subset \mathbb{R}$, can be represented as the sum of cosine and sine functions of various frequencies with various amplitudes:
$$ f(x) = \sum_{m=0}^{\infty} A_{m}\cos \left( \frac{m \pi x}{L} \right) + B_{m}\sin \left( \frac{m \pi x}{L} \right) $$
This infinite sum is famously known as the Fourier series \cite{fourier_series}. In the Kolmogorov-Arnold Network, we will approximate each trainable function $\phi_{j,k}^{(i)}$ using Fourier series with a finite number of terms. Define $F$ as the number of terms to be used, then we have the following
$$ \phi_{j,k}^{(i)}(x) = \sum_{m=1}^{F} A_{j,k}^{(i)}(m)\cos \left( \omega_{j,k}^{(i)}(m) x\right) + B_{j,k}^{(i)}(m)\sin \left( \omega_{j,k}^{(i)}(m) x\right). $$
Thus, training the functions $\phi_{j,k}^{(i)}$s is equivalent as training the cosine amplitudes $A_{j,k}^{(i)}(m)$, the sine amplitudes $B_{j,k}^{(i)}(m)$, and the frequencies $\omega_{j,k}^{(i)}(m)$, for $m=1,\hdots,F$. 


\section{SIRD Model}
\label{sec:sird_definition}

\subsection{Model Definition}
\noindent The following is our SIRD compartmental population model:
\begin{align} 
\frac{dS}{dt} &= -\beta(t) SI  \label{sird_model_S} \\
\frac{dI}{dt} &= \beta(t) SI  - \gamma (t) I - \mu (t)I \label{sird_model_I} \\
\frac{dR}{dt} &= \gamma (t)I  \label{sird_model_R} \\ 
\frac{dD}{dt} &= \mu (t) I  \label{sird_model_D} 
\end{align}
The variables involved are the susceptible population ($S$), the currently infected population ($I$), the recovered population ($R$), and the deceased population ($D$). The parameter $\beta(t)$ represents the transmission rate at time $t$, the parameter $\gamma(t)$ represents the recovery rate of transfer from the currently infected population to the recovered compartment $R$ at time $t$. The last parameter $\mu(t)$ is the mortality rate of transfer from the currently infected population to the deceased compartment $D$ at time $t$. The model operates under the assumption that there is no birth and deceased from factors other than COVID-19, and there is no reinfection.

If we want to model the epidemic dynamics in terms of relative proportion, instead of the population number itself, we can let $N_{p}$ be the initial number of population and define the variables: $S^{rel}=\frac{S}{N_{p}},I^{rel}=\frac{I}{N_{p}},R^{rel}=\frac{R}{N_{p}},D^{rel}=\frac{D}{N_{p}}$. Therefore, our SIRD system (\ref{sird_model_S})-(\ref{sird_model_D}) can be transformed into the following:
\begin{align} 
\frac{dS^{rel}}{dt} &= -\beta^{rel}(t) S^{rel} I^{rel}  \label{sird_model_S_relative} \\
\frac{dI^{rel}}{dt} &= \beta^{rel}(t) S^{rel}I^{rel}  - \gamma (t) I^{rel} - \mu (t)I^{rel} \label{sird_model_I_relative} \\
\frac{dR^{rel}}{dt} &= \gamma (t)I^{rel}  \label{sird_model_R_relative} \\ 
\frac{dD^{rel}}{dt} &= \mu (t) I^{rel}  \label{sird_model_D_relative}, 
\end{align}
where $\beta^{rel} = N_{p}\beta$. Note that only the relative transmission rate is $\beta^{rel}(t)$ in the relative proportion system that are different than the actual transmission rate $\beta(t)$ of the actual population system. The other two rates $\gamma(t), \mu(t)$ are the same for both systems. In the next two sections (Section \ref{sec:data_preparation} and \ref{sec:data_preparation_NN}), we explain how we obtain and process the COVID-19 data, before using them to estimate the time-dependent rates.

\section{Processing COVID-19 Data} \label{sec:data_preparation}

\noindent We consider the COVID-19 dataset from several countries: Indonesia, Singapore, Malaysia. For each country, we denote the estimated initial number of population on 1 February 2020 as $N_{p}$, collected from United Nation's World Population Prospects 2024 \cite{un_wpp}. Let the number of data points (time points) for each country is denoted as $n_{t}$. We now explain how to construct the actual solution for $S(t),I(t),R(t),D(t)$ at discrete time points, using COVID-19 data. Let $t_{n}= n \Delta t, \:\:\: n=0,\hdots,n_{t}$ be the discrete time points with $\Delta t > 0$ small enough. In this paper, we use $\Delta t = 1$. For simplicity, we write $S_{n}=S(t_{n}), I_{n}=I(t_{n}), R_{n}=R(t_{n}), D_{n}=D(t_{n})$ for each $n=0,\ldots,n_{t}$. We collect the data of cumulative number of cases, cumulative number of deceased, and the number of currently infected population (active cases) from Worldometer \cite{worldometers}. The data for the susceptible compartment $S_{n}$ is approximated as follows:
\begin{equation}\label{S_data_formula}
S_{n} = N_{p} - \left( \text{Cumulative number of cases at time } t=t_{n} \right).  
\end{equation}
The data for the recovered compartment $R_{n}$ can be approximated as follows:
\begin{equation}\label{R_data_formula}
R_{n} = N_{p} - \left( S_{n} + I_{n} + D_{n} \right).
\end{equation}
 
\section{Training Data}\label{sec:data_preparation_NN}

\noindent As we have explained in Section \ref{sec:intro}, each of the three neural networks will have either 8 or 4 input variables. To work with neural networks, all the data that will be used as input values for the neural networks should be normalized. Normalization of input values can improve optimization (or minimization) performance. Here, we use the standard $z$-score method so that the data from each variable will have mean equal to 0 and standard deviation equal to 1. We introduce the following notation:
\begin{align*} 
\Delta S^{rel}(t) &= S^{rel}(t + \Delta t)-S^{rel}(t), \\ 
\Delta I^{rel}(t) &= I^{rel}(t + \Delta t) - I^{rel}(t), \\ 
\Delta R^{rel}(t) &= R^{rel}(t + \Delta t) - R^{rel}(t), \\
\Delta D^{rel}(t) &= D^{rel}(t + \Delta t) - D^{rel}(t). 
\end{align*}
To estimate COVID-19 parameters at a particular time $t$, the input values can include 4 normalized epidemic variables (we will normalize $S^{rel}(t), I^{rel}(t), R^{rel}(t), D^{rel}(t)$), and 4 normalized gradient variables (we will normalize $\Delta S^{rel}(t), \Delta I^{rel}(t), \Delta R^{rel}(t), \Delta D^{rel}(t)$). In this section, we explain more about how to normalize these variables and prepare training data. 

\subsection{Normalized COVID-19 Variables}
\noindent In this subsection we construct the 4 normalized COVID-19 variables. We first divide all variables with the initial population $N_{p}$ to obtain: $S_{n}^{rel}=\frac{S_{n}}{N_{p}}, I_{n}^{rel}=\frac{I_{n}}{N_{p}}, R_{n}^{rel}=\frac{R_{n}}{N_{p}}, D_{n}^{rel}=\frac{D_{n}}{N_{p}}$. Subsequently, let $\tilde{S_{n}},\tilde{I_{n}},\tilde{R_{n}},\tilde{D_{n}}$ be the normalized data point at time point $t_{n}$. We normalize the data of each compartment differently using standard $z$-score as follows:
\begin{align*}
\tilde{S_{n}} &= \frac{S_{n}^{rel}-Mean\left( \{S_{i}^{rel}\}_{i=0}^{i=n_{t}-1} \right)}{Std\left( \{S_{i}^{rel}\}_{i=0}^{i=n_{t}-1} \right)}, \:\: \\
\tilde{I_{n}} &= \frac{I_{n}^{rel}-Mean\left( \{I_{i}^{rel}\}_{i=0}^{i=n_{t}-1} \right)}{Std\left( \{I_{i}^{rel}\}_{i=0}^{i=n_{t}-1} \right)}, \:\: \\
\tilde{R_{n}} &= \frac{R_{n}^{rel}-Mean\left( \{R_{i}^{rel}\}_{i=0}^{i=n_{t}-1} \right)}{Std\left( \{R_{i}^{rel}\}_{i=0}^{i=n_{t}-1} \right)}, \:\: \\
\tilde{D_{n}} &= \frac{D_{n}^{rel}-Mean\left( \{D_{i}^{rel}\}_{i=0}^{i=n_{t}-1} \right)}{Std\left( \{D_{i}^{rel}\}_{i=0}^{i=n_{t}-1} \right)}, \:\:
\end{align*}
where the $Mean(\cdot)$ and $Std(\cdot)$ are the arithmetic mean and standard deviation functions, respectively. In general, we may also define the normalized values at arbitrary time points $t \ge 0$: 
\begin{align*}
\tilde{S}(t) &= \frac{S^{rel}(t) - Mean\left( \{S_{i}^{rel}\}_{i=0}^{i=n_{t}-1} \right)}{Std\left( \{S_{i}^{rel}\}_{i=0}^{i=n_{t}-1} \right)}, \:\: \\
\tilde{I}(t) &= \frac{I^{rel}(t)-Mean\left( \{I_{i}^{rel}\}_{i=0}^{i=n_{t}-1} \right)}{Std\left( \{I_{i}^{rel}\}_{i=0}^{i=n_{t}-1} \right)}, \:\: \\
\tilde{R}(t) &= \frac{R^{rel}(t)-Mean\left( \{R_{i}^{rel}\}_{i=0}^{i=n_{t}-1} \right)}{Std\left( \{R_{i}^{rel}\}_{i=0}^{i=n_{t}-1} \right)}, \:\: \\
\tilde{D}(t) &= \frac{D^{rel}(t)-Mean\left( \{D_{i}^{rel}\}_{i=0}^{i=n_{t}-1} \right)}{Std\left( \{D_{i}^{rel}\}_{i=0}^{i=n_{t}-1} \right)}, \:\:
\end{align*}
where $S^{rel}(t) = \frac{S(t)}{N_{p}}$, $I^{rel}(t) = \frac{I(t)}{N_{p}}$, $R^{rel}(t) = \frac{R(t)}{N_{p}}$, $D^{rel}(t) = \frac{D(t)}{N_{p}}$.

\subsection{Normalized Gradients}
\noindent We have explained that the neural networks' input layers can have additional 4 variables of normalized numerical gradients. The numerical gradients are defined as
\begin{align*}
    \Delta S_{n}^{rel} &= S_{n+1}^{rel}-S_{n}^{rel}, \:\:
    \Delta I_{n}^{rel} = I_{n+1}^{rel}-I_{n}^{rel} \\
    \Delta R_{n}^{rel} &= R_{n+1}^{rel}-R_{n}^{rel}, \:\:
    \Delta D_{n}^{rel} = D_{n+1}^{rel}-D_{n}^{rel} 
\end{align*}
For each compartment, we then normalize these values according to the standard $z$-score method:
\begin{align*}
\Delta \tilde{S}_{n} &= \frac{\Delta S_{n}^{rel}-Mean\left( \{\Delta S_{i}^{rel}\}_{i=0}^{i=n_{t}-1} \right)}{Std\left( \{\Delta S_{i}^{rel}\}_{i=0}^{i=n_{t}-1} \right)}, \:\: \\
\Delta \tilde{I}_{n} &= \frac{\Delta I_{n}^{rel} - Mean\left( \{\Delta I_{i}^{rel}\}_{i=0}^{i=n_{t}-1} \right)}{Std\left( \{\Delta I_{i}^{rel}\}_{i=0}^{i=n_{t}-1} \right)}, \:\: \\
\Delta \tilde{R}_{n} &= \frac{\Delta R_{n}^{rel}-Mean\left( \{\Delta R_{i}^{rel}\}_{i=0}^{i=n_{t}-1} \right)}{Std\left( \{\Delta R_{i}^{rel}\}_{i=0}^{i=n_{t}-1} \right)}, \:\: \\
\Delta \tilde{D}_{n} &= \frac{\Delta D_{n}^{rel}-Mean\left( \{\Delta D_{i}^{rel}\}_{i=0}^{i=n_{t}-1} \right)}{Std\left( \{\Delta D_{i}^{rel}\}_{i=0}^{i=n_{t}-1} \right)}.
\end{align*}
In general, we may also define the normalized gradients at arbitrary time points $t \ge 0$. The differences at arbitrary time $t$ are
\begin{align*}
    \Delta S^{rel}(t) &= S^{rel}(t+\Delta t)-S^{rel}(t), \:\:
    \Delta I^{rel}(t) = I^{rel}(t+\Delta t)-I^{rel}(t),  \\
    \Delta R^{rel}(t) &= R^{rel}(t+\Delta t)-R^{rel}(t),  \:\:
    \Delta D^{rel}(t) = D^{rel}(t+\Delta t)-D^{rel}(t). 
\end{align*}
In this paper, we use $\Delta t = 1$. Therefore, the normalized gradients at arbitrary time point $t$ are 
\begin{align*}
\Delta \tilde{S}(t) &= \frac{\Delta S^{rel}(t) - Mean\left( \{\Delta S_{n}^{rel}\}_{i=0}^{i=n_{t}-1} \right)}{Std\left( \{\Delta S_{i}^{rel}\}_{i=0}^{i=n_{t}-1} \right)}, \:\: \\
\Delta \tilde{I}(t) &= \frac{\Delta I^{rel}(t) - Mean\left( \{\Delta I_{i}^{rel}\}_{i=0}^{i=n_{t}-1} \right)}{Std\left( \{\Delta I_{i}^{rel}\}_{i=0}^{i=n_{t}-1} \right)}, \:\: \\
\Delta \tilde{R}(t) &= \frac{\Delta R^{rel}(t) -Mean\left( \{\Delta R_{i}^{rel}\}_{i=0}^{i=n_{t}-1} \right)}{Std\left( \{\Delta R_{i}^{rel}\}_{i=0}^{i=n_{t}-1} \right)}, \:\: \\
\Delta \tilde{D}(t) &= \frac{\Delta D^{rel}(t) -Mean\left( \{\Delta D_{i}^{rel}\}_{i=0}^{i=n_{t}-1} \right)}{Std\left( \{\Delta D_{i}^{rel}\}_{i=0}^{i=n_{t}-1} \right)}.
\end{align*}

\section{KAN-F COVID-19 Parameter Model} \label{sec:kan_fourier_specific}
\noindent According to Model A, we define $\widehat{\beta_{a}}: \mathbb{R}^{8} \rightarrow (0,1)$, $\widehat{\gamma_{a}}: \mathbb{R}^{8} \rightarrow (0,1)$, and $\widehat{\mu_{a}}: \mathbb{R}^{8} \rightarrow (0,1)$ as a set of models that approximate the true transmission rate $\beta$, true recovery rate $\gamma$, and true mortality rate $\mu$, respectively. The input for each of $\widehat{\beta_{a}}$, $\widehat{\gamma_{a}}$, and $\widehat{\mu_{a}}$ is the vector $\mathbf{x}[t] = [\tilde{S}(t), \tilde{I}(t), \tilde{R}(t), \tilde{D}(t), \Delta \tilde{S}(t), \Delta \tilde{I}(t), \Delta \tilde{R}(t), \Delta \tilde{D}(t)]$, using $\Delta t = 1$. Similarly, according to Model B, we define $\widehat{\beta_{b}}: \mathbb{R}^{4} \rightarrow (0,1)$, $\widehat{\gamma_{b}}: \mathbb{R}^{4} \rightarrow (0,1)$, and $\widehat{\mu_{b}}: \mathbb{R}^{4} \rightarrow (0,1)$ as another set of models that approximate the true transmission rate $\beta$, true recovery rate $\gamma$, and true mortality rate $\mu$, respectively. The input for each of $\widehat{\beta_{b}}$, $\widehat{\gamma_{b}}$, and $\widehat{\mu_{b}}$ is the vector $\mathbf{x}[t] = [\tilde{S}(t), \tilde{I}(t), \tilde{R}(t), \tilde{D}(t)]$. Since transmission, recovery, and mortality rates should not have values of greater than 1, we improvise by mapping the final output of the network to the sigmoid function: $Sigmoid(x)=\frac{e^{x}}{1+e^{x}}$. Hence, we may write $\widehat{\beta} = Sigmoid(\widehat{f}_{\beta}(\mathbf{x}[t]))$, $\widehat{\gamma} = Sigmoid(\widehat{f}_{\gamma}(\mathbf{x}[t]))$, $\widehat{\mu} = Sigmoid(\widehat{f}_{\mu}(\mathbf{x}[t]))$, where the $\widehat{f}_{\beta},\widehat{f}_{\gamma},\widehat{f}_{\mu}$ are KAN-F networks. We set $\widehat{f}_{\beta}$, $\widehat{f}_{\gamma}$, and $\widehat{f}_{\mu}$ as three different KAN-F networks with the same layer-sequence $[8,17,35,1]$ (for Model A) and $[4,9,19,1]$ (for Model B). For Model A, note that the last hidden layer has 35 neurons, which is twice the number of neurons in the first hidden layer, 17, and then plus one. The same pattern of layer-sequence also applies for Model B. The number of Fourier terms in every Fourier series or activation function is set to $F=30$. We would like to optimize these networks by minimizing an objective loss function $\mathcal{L}$ defined in the following section.

\section{Multi-Objective Optimization} 
\label{sec:loss_function}
\noindent In this section, we define the objective loss function that must be minimized in order to obtain the optimum Fourier amplitudes and frequencies of the KAN-F networks. The optimum amplitudes and frequencies should make the networks to be able to estimate transmission, recovery, and mortality rates accurately. The process of this minimization is also called neural network training. For brevity, let $\mathbf{x}_{n} = \mathbf{x}[t_{n}]$.  Let $\vec{\theta}_{\beta}$, $\vec{\theta}_{\gamma}$, and $\vec{\theta}_{\mu}$ be the sets that contain only the trainable NN parameters (i.e. the amplitudes and frequencies of all Fourier series) for each neural network models $\widehat{\beta}$, $\widehat{\gamma}$, and $\widehat{\mu}$, respectively. We write $\widehat{\beta}(\mathbf{x}_{n}; \vec{\theta}_{\beta})$ for the evaluation of the KAN-F $\widehat{\beta}$ at vector $\mathbf{x}_{n}$ using $\vec{\theta}_{\beta}$ as the NN parameters. We also define $\widehat{\gamma}(\mathbf{x}_{n}; \vec{\theta}_{\gamma})$, and $\widehat{\mu}(\mathbf{x}_{n}; \vec{\theta}_{\mu})$ the same way. Let $\vec{\theta}$ be the vector of NN parameters from all the 3 network functions combined. In the next subsection, we first describe the Runge-Kutta fourth-order method in the context of our SIRD model.

\begin{figure}
    \centering 
\makebox[\textwidth]{
\includegraphics[width=16cm]{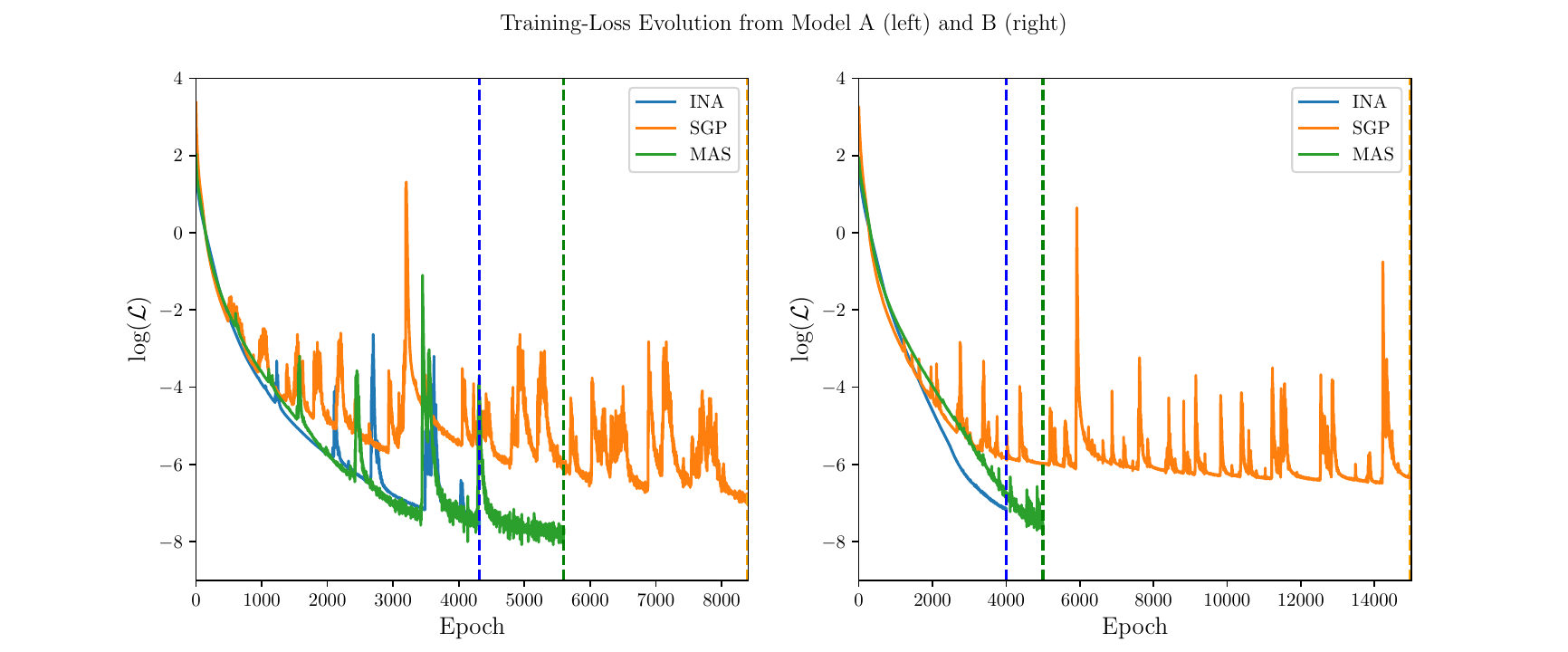}}
    \caption{The values of $\log(\mathcal{L})$ at different epochs. The dashed lines represent the end of the training process for each country. }
\label{fig:log_loss}
\end{figure}

\subsection{Runge-Kutta Numerical Scheme} \label{sec:runge_kutta}
\noindent If we replace the three actual epidemic rates in the SIRD system (\ref{sird_model_S_relative})-(\ref{sird_model_D_relative}) with the 3 network functions $\widehat{\beta}, \widehat{\gamma}, \widehat{\mu}$, then (\ref{sird_model_S_relative})-(\ref{sird_model_D_relative}) can be rewritten as $\frac{dY(t)}{dt} = F(Y(t))$, where
\begin{align*}
    Y(t) = \begin{pmatrix}
        S^{rel}(t) \\ I^{rel}(t) \\ R^{rel}(t) \\ D^{rel}(t)
    \end{pmatrix}, \:\:\: 
    F(Y(t)) &= \begin{pmatrix}
    -\widehat{\beta}(\mathbf{x}[t])S^{rel}I^{rel}  \\
    \widehat{\beta}(\mathbf{x}[t])S^{rel}I^{rel}  -\widehat{\gamma}(\mathbf{x}[t])I^{rel} - \widehat{\mu}(\mathbf{x}[t])I^{rel}  \\
   \widehat{\gamma}(\mathbf{x}[t])I^{rel} \\
   \widehat{\mu}(\mathbf{x}[t])I^{rel}
    \end{pmatrix}
\end{align*}
If $Y(t_{n})$ is known, then the Runge-Kutta fourth-order scheme \cite{runge_kutta} approximates $Y(t_{n+1})$ by first calculating the following:
\begin{align*}
K_{1} &= \Delta t F(Y(t_{n})) = \Delta t, \:\:
K_{2} = \Delta t F \left( Y(t_{n}) + \frac{K_{1}}{2} \right),\\
K_{3} &=  \Delta t F \left( Y(t_{n}) + \frac{K_{2}}{2} \right), \:\:
K_{4} =  \Delta t F \left( Y(t_{n}) + K_{3} \right).
\end{align*}
Note that $K_{1},K_{2},K_{3},K_{4}$ are all vectors with 4 elements. Subsequently, these values are then used to approximate $Y(t_{n+1})$ as follows:
\begin{equation}\label{RK_approx}
    Y(t_{n+1}) \approx Y(t_{n}) + \frac{1}{6}K_{1} + \frac{1}{3}K_{2} + \frac{1}{3}K_{3} + \frac{1}{6}K_{4}.
\end{equation}
Let $\vec{r}_{n}(\vec{\theta})$ be the vector containing the three network functions evaluated at time $t_{n}$ under the vector of NN parameters $\vec{\theta}$. Thus, we write $\vec{r}_{n}(\vec{\theta})$ as follows:
\begin{equation*}
    \vec{r}_{n}(\vec{\theta}) = [\widehat{\beta}(\mathbf{x}_{n};\vec{\theta}_{\beta}),\widehat{\gamma}(\mathbf{x}_{n};\vec{\theta}_{\gamma}), \widehat{\mu}(\mathbf{x}_{n};\vec{\theta}_{\mu})]
\end{equation*}
We define $RK(t_{n};\vec{r}_{n}(\vec{\theta}))$ as the right hand side of (\ref{RK_approx}) computed using rate values $\vec{r}_{n}(\vec{\theta})$, and we will call from the 1st to the 4th element of it as 
\begin{equation}
    RK(t_{n};\vec{r}_{n}(\vec{\theta})) =
    [RK(S_{n}^{rel};\vec{r}_{n}(\vec{\theta})),RK(I_{n}^{rel};\vec{r}_{n}(\vec{\theta})),RK(R_{n}^{rel};\vec{r}_{n}(\vec{\theta})),RK(D_{n}^{rel};\vec{r}_{n}(\vec{\theta}))],
\end{equation}
so that the following relation holds:
\begin{equation}
Y(t_{n+1}) \approx RK(t_{n};\vec{r}_{n}(\vec{\theta})) = \begin{pmatrix}
RK(S_{n}^{rel};\vec{r}_{n}(\vec{\theta})) \\ 
RK(I_{n}^{rel};\vec{r}_{n}(\vec{\theta})) \\ 
RK(R_{n}^{rel};\vec{r}_{n}(\vec{\theta})) \\
RK(D_{n}^{rel};\vec{r}_{n}(\vec{\theta}))   
\end{pmatrix} = Y(t_{n}) + \frac{1}{6}K_{1} + \frac{1}{3}K_{2} + \frac{1}{3}K_{3} + \frac{1}{6}K_{4}, \label{RK_notation}
\end{equation}
for $0 \le n \le n_{t}-1$.

\subsection{Loss Function Definition}
Let the interval $[t_{0},t_{n_{t}}]$ be a particular time period where we want to estimate the rates. Given a subset of relative proportion data values of $S_{n}^{rel},I_{n}^{rel},R_{n}^{rel},D_{n}^{rel}$, for $0 \le n \le n_{t}$, which are considered as a solution to the SIRD model (\ref{sird_model_S_relative})-(\ref{sird_model_D_relative}) at discrete time points, we want to find the Fourier amplitudes of all the network functions $\widehat{\beta},\widehat{\gamma},\widehat{\mu}$ such that the following mean of squared errors are minimized: 
\begin{align}
\mathcal{L}_{S}(\vec{\theta}) &= \Omega_{S} \sqrt{ Mean\left( \left\{ \left[ (S_{i+1}^{rel} - RK(S_{i}^{rel};\vec{r}_{i}(\vec{\theta})))^{2} \right] \right\}_{i=0}^{n_{t}-1}  \right) }, \label{loss_S}\\
\mathcal{L}_{I}(\vec{\theta}) &= \Omega_{I} \sqrt{ Mean\left( \left\{ \left[ (I_{i+1}^{rel} - RK(I_{i}^{rel};\vec{r}_{i}(\vec{\theta})))^{2} \right]\right\}_{i=0}^{n_{t}-1}  \right) }, \label{loss_I}\\
\mathcal{L}_{R}(\vec{\theta}) &= \Omega_{R} \sqrt{ Mean\left( \left\{ \left[ (R_{i+1}^{rel} - RK(R_{i}^{rel};\vec{r}_{i}(\vec{\theta})))^{2} \right] \right\}_{i=0}^{n_{t}-1}  \right) }, \label{loss_R}\\
\mathcal{L}_{D}(\vec{\theta}) &= \Omega_{D}\sqrt{ Mean\left( \left\{ \left[ (D_{i+1}^{rel} - RK(D_{i}^{rel};\vec{r}_{i}(\vec{\theta})))^{2} \right]\right\}_{i=0}^{n_{t}-1}  \right) }, \label{loss_D}
\end{align}
with $\Delta t = 1$, and with $\Omega_{S},\Omega_{I},\Omega_{R},\Omega_{D}$ as predetermined loss balancing factors. These are the root-mean-of-squared errors (RMSE) between the exact solution against the Runge-Kutta numerical solution for all compartments at time points $\{t_{n}\}_{n=1}^{n_{t}+1}$. This approach was also adopted in \cite{xiao_fourier_mapping}. The combined total loss function that we want to minimize is
\begin{align}  \label{basic_loss_function}
\mathcal{L}(\vec{\theta}) &= \mathcal{L}_{S}(\vec{\theta})  +  \mathcal{L}_{I}(\vec{\theta}) + \mathcal{L}_{R}(\vec{\theta}) + \mathcal{L}_{D}(\vec{\theta})
\end{align}
We can minimize (\ref{basic_loss_function}) by a gradient descent algorithm. The computation in this algorithm makes use of gradients $\nabla \mathcal{L}$ with respect to multitudinous neural network parameters in $\vec{\theta}$. We use the ADAM (Adaptive Moment Estimation) Optimizer, which is a type of gradient descent algorithm \cite{kingma}. We use Python and its Pytorch module \cite{pytorch_site} to use the ADAM Optimizer.

\section{Results} \label{sec:results}

\begin{figure}
    \centering 
\makebox[\textwidth]{
\includegraphics[width=17cm]{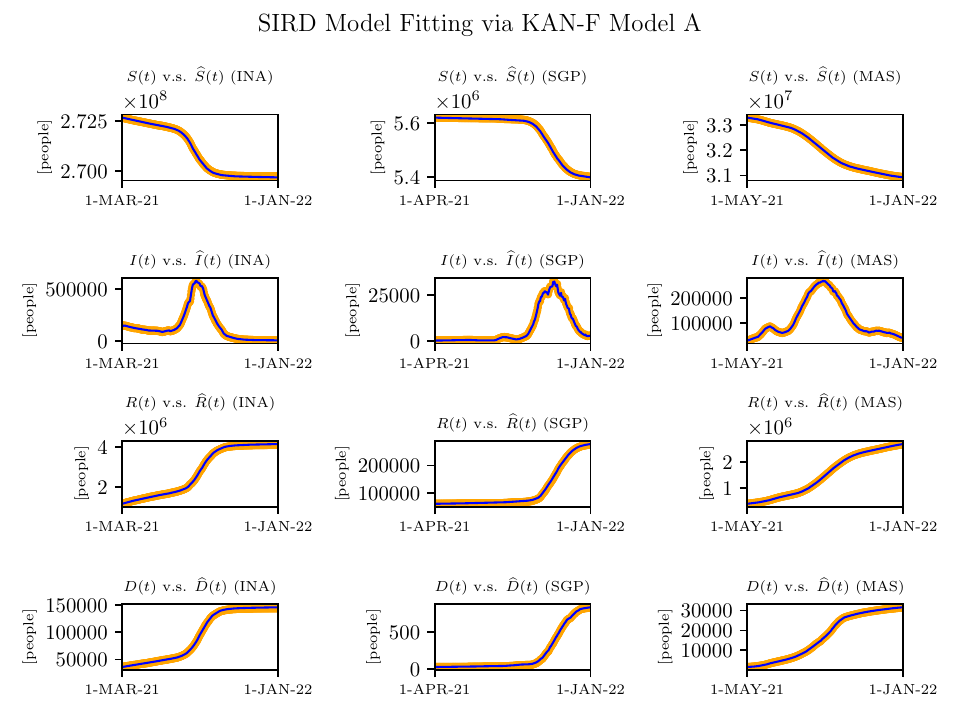}}
    \caption{The model-fitting between Runge-Kutta's numerical solution of our SIRD model (blue) against the actual COVID-19 data (orange). The first column is Indonesia (INA), second column is Singapore (SGP), and third column is Malaysia (MAS). This result uses estimated rates from Model A.}
\label{fig:model_fitting_all}
\end{figure}

\begin{figure}
    \centering 
\makebox[\textwidth]{
\includegraphics[width=17cm]{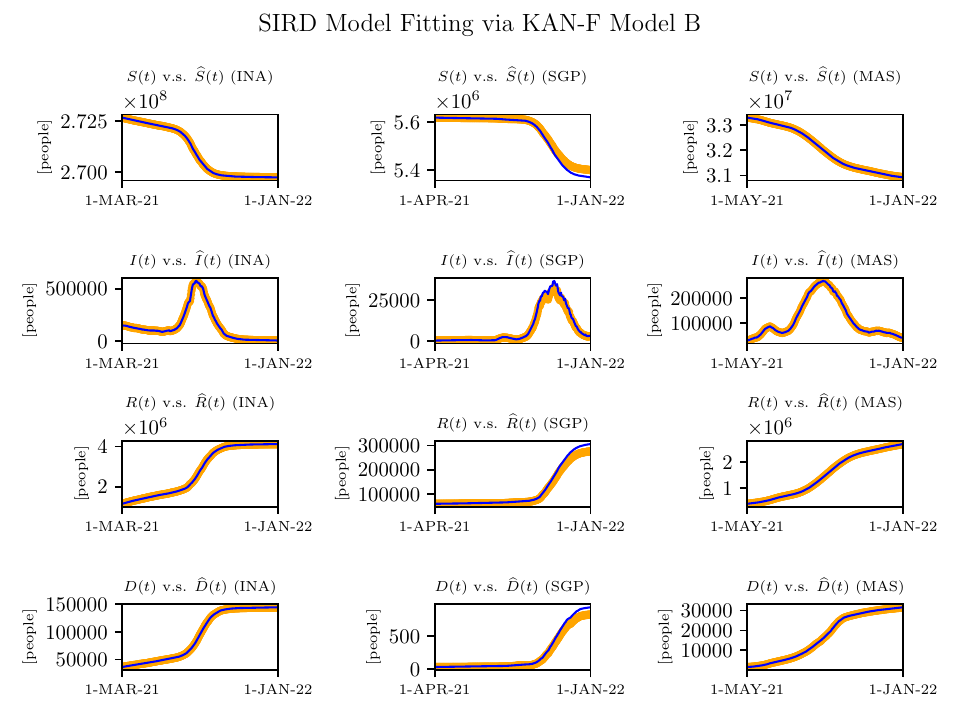}}
    \caption{The model-fitting between Runge-Kutta's numerical solution of our SIRD model (blue) against the actual COVID-19 data (orange). The first column is Indonesia (INA), second column is Singapore (SGP), and third column is Malaysia (MAS). This result uses estimated rates from Model B.}
\label{fig:model_fitting_all_PREDICTOR}
\end{figure}

\noindent We estimate three parameters of the COVID-19 model, these are the time-dependent transmission rate $\beta(t)$, recovery rate $\gamma(t)$, and mortality rate $\mu(t)$, during the phase in which the SARS-CoV-2 Delta variant was predominant. Recall that these time-dependent rates are in the context of SIRD compartmental differential equations (\ref{sird_model_S}-\ref{sird_model_D}). We consider three countries in South-East Asia: Indonesia, Singapore, and Malaysia.  We use the KAN-F layer-sequence of $[8, 17, 35, 1]$ (for Model A) and $[4, 9, 19, 1]$ (for Model B) for all countries. Therefore, each network comprises 8 or 4 input neurons, 2 hidden layers, and 1 output neuron. For all countries, we use $F=30$, which means there are 30 cosine terms and 30 sine terms for each learnable activation function represented by Fourier series. Our final estimated rates are shown in Figure \ref{fig:estimated_rates_delta} and \ref{fig:estimated_rates_delta_PREDICTOR}. The accuracy of COVID-19 parameters estimation is measured by the model-fitting between the numerical solution of the SIRD model (\ref{sird_model_S}-\ref{sird_model_D}) and the real-world data. Overall, we have decent results for all countries: the model-fitting plots are shown in Figure \ref{fig:model_fitting_all} and \ref{fig:model_fitting_all_PREDICTOR} and the RMSE measures of these fitting are summarized in Table \ref{table:model_fitting_all} and \ref{table:model_fitting_all_PREDICTOR}. Visually, the model-fitting plots in Figure \ref{fig:model_fitting_all} show that our estimated rates are accurate enough. However, in the Singapore case for Model B, it is difficult for the model to appropriately fit the data even after 15000 epochs. Numerically, the RMSE results shown in Table \ref{table:model_fitting_all} and \ref{table:model_fitting_all_PREDICTOR} are very small relative to the peaks of the susceptible, infective, recovered, and deceased population. Moreover, our methods are comparatively efficient: the number of epochs needed to train our KAN-F models for all countries are less than 10000. The Indonesia case requires 4316 epochs for Model A and 4000 epochs for Model B. The Singapore case requires 8408 epochs for Model A, although it is difficult to attain good fitting under Model B. The Malaysia case requires 5600 epochs for Model A and 5000 epochs for Model B. The training dynamics for all countries are plotted in Figure \ref{fig:log_loss} as the values of $\log(\mathcal{L})$ at different epochs. For one country, the KAN-F training process is independent of the other two countries. Next, we discuss the implementation details for each country's estimation (or KAN-F training process).

\subsection{Indonesia}
\noindent For Indonesia, SARS-CoV-2 Delta variant began to appear around 1 March 2021 and declined in influence around 1 January 2022 \cite{covid_variants}. The estimated initial population on 1 February 2020 for Indonesia is $N_{p}=274$ million \cite{un_wpp}. The loss balancing factors are $\Omega_{S}=1$, $\Omega_{I}=500$, $\Omega_{R}=500$, $\Omega_{D}=20000$. The learning rate for the ADAM optimizer is set to $2 \times 10^{-5}$. We require the training process to stop once the loss value (\ref{basic_loss_function}) is lower than $5\times 10^{-4}$. 

\subsection{Singapore}
\noindent For Singapore, SARS-CoV-2 Delta variant began to appear around 1 April 2021 and declined in influence around 1 January 2022 \cite{covid_variants}. The estimated initial population on 1 February 2020 for Singapore is $N_{p}=5.68$ million \cite{un_wpp}. The loss balancing factors are $\Omega_{S}=1$, $\Omega_{I}=700$, $\Omega_{R}=60$, $\Omega_{D}=30000$. The learning rate for the ADAM optimizer is set to $5 \times 10^{-5}$.

\subsection{Malaysia}
\noindent For Malaysia, SARS-CoV-2 Delta variant began to appear around 1 May 2021 and declined in influence around 1 January 2022 \cite{covid_variants}. The estimated initial population on 1 February 2020 for Malaysia is $N_{p}=33.697$ million \cite{un_wpp}. The loss balancing factors are $\Omega_{S}=1$, $\Omega_{I}=280$, $\Omega_{R}=50$, $\Omega_{D}=4000$. The learning rate for the ADAM optimizer is set to $2 \times 10^{-5}$.

\begin{table}
\centering
\begin{tabular} {|c|c|c|c|c|c|c|c|c| }
 \hline
 &  \multicolumn{2}{|c|}{RMSE $S$}  & \multicolumn{2}{|c|}{RMSE $I$}  & \multicolumn{2}{|c|}{RMSE $R$}  & \multicolumn{2}{|c|}{RMSE $D$}   \\ \hline
 & Val & Val/max & Val & Val/max & Val & Val/max & Val & Val/max \\ 
 \hline
INA & 21068 & 0.0077\% & 2338  & 0.41\%  & 18920 & 0.46\% & 844 & 0.59\%  \\
\hline
SGP & 2115 & 0.038\% & 130  & 0.4\%  & 2172  & 0.79\% & 6 & 0.69\%  \\
\hline
MAS & 720 & 0.0022\% & 176  & 0.066\%  & 667  & 0.025\% & 41 & 0.13\%  \\
\hline
\end{tabular}
\caption{The Root-Mean-Squared-Error (RMSE) of the model fitting via KAN-F Model A. Val is the RMSE, while Val/max is the RMSE divided by the highest number in the corresponding compartment. For example: the peak of Indonesia's number of active case data ($I$) was 574135 people, so the RMSE for $I$ (which is 2338) is divided by 574135 to obtain $0.41$\%. } \label{table:model_fitting_all}
\end{table}

\begin{table}
\centering
\begin{tabular} {|c|c|c|c|c|c|c|c|c| }
 \hline
 &  \multicolumn{2}{|c|}{RMSE $S$}  & \multicolumn{2}{|c|}{RMSE $I$}  & \multicolumn{2}{|c|}{RMSE $R$}  & \multicolumn{2}{|c|}{RMSE $D$}   \\ \hline
 & Val & Val/max & Val & Val/max & Val & Val/max & Val & Val/max \\ 
 \hline
INA & 2247 & 0.0008\% & 244  & 0.04\%  & 1837 & 0.045\% & 242 & 0.17\%  \\
\hline
SGP & 14240 & 0.25\% & 1582 	& 4.82\%  & 13135  & 4.76\% & 48 & 5.76\%  \\
\hline
MAS & 4190 & 0.013\% & 416  & 0.15\%  & 3863  & 0.144\% & 80 & 0.25\%  \\
\hline
\end{tabular}
\caption{The Root-Mean-Squared-Error (RMSE) of the model fitting via KAN-F Model B. Val is the RMSE, while Val/max is the RMSE divided by the highest number in the corresponding compartment. For example: the peak of Indonesia's number of active case data ($I$) was 574135 people, so the RMSE for $I$ (which is 244) is divided by 574135 to obtain $0.04$\%. } \label{table:model_fitting_all_PREDICTOR}
\end{table}

\section{Forecasting} \label{sec:forecast}
\begin{figure}[H]
    \centering 
\makebox[\textwidth]{
\includegraphics[width=17cm]{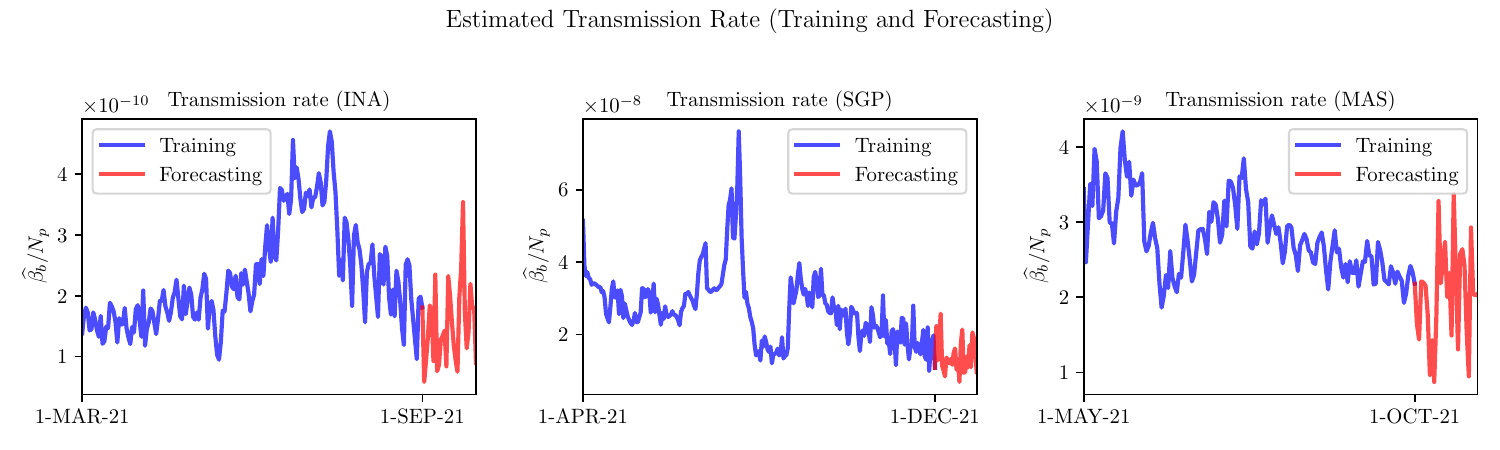}}
    \caption{The estimated transmission rates for all three countries. The blue line indicates transmission rates during training period, while the red line is the rates during forecast period. In this plot, the transmission rate during forecast period is already scaled by the factor of 17\% (Indonesia) and 30 \% (Singapore and Malaysia).}
\label{fig:estimated_rates_pdf_FORECAST}
\end{figure}

\noindent When we have the data of the number of susceptibles, active cases, recovered, and deceased at a particular time $t$, we would like to estimate what are the transmission, recovery, and mortality rates at the same time $t$, which will affect the dynamics up to time $t+\Delta t$. This is what we have done using Model B. In this section, we will predict the transmission rates outside of the training period and training data. After we have trained our KAN-F functions of Model B inside the time interval [$t_{start}$, $t_{end}$] (training period), we would like to see the output of the KAN-F functions inside the time interval $[t_{end}, t_{end}+29]$. We call this time period a forecast period. The output of the KAN-F functions inside the forecast period will be used to predict the cumulative number of cases of COVID-19 for the next 30 days. We train $\widehat{\beta}_{b}$, $\widehat{\gamma}_{b}$, and $\widehat{\mu}_{b}$ inside the time interval [$t_{start}$, $t_{end}$]. Subsequently, we use the output of the trained $\widehat{\beta}_{b}$ for each day in \{$t=t_{end}$, $\hdots$, $t=t_{end}+29$\}. However, we found that the raw output of $\widehat{\beta}_{b}$ inside the forecast interval [$t_{end}$, $t_{end}+29$] is an overestimation and not in proportion with the estimated rates during the training period [$t_{start}$, $t_{end}$]. To make them more accurate, we perform an adjustment: we scale them so that they become proportional to the estimated rates during the training period and resulting in decent predictions for all three countries.  We found that multiplying the output of $\widehat{\beta}_{b}$ at the discrete points \{$t_{end}$, $\hdots$, $t_{end}+29$\} by the factors 17\% (for Indonesia) and 30\% (for Singapore and Malaysia), gives significantly better prediction than using the raw original output. This approach is similar to the study in \cite{covid19_rt_factor} for predicting COVID-19 epidemic trends using a neural network. The estimated rates during training and forecast period are plotted in Figure \ref{fig:estimated_rates_pdf_FORECAST}). We use the same KAN-F Model B and the same optimization setup as in the previous section. For Indonesia, the training period is between 1 March 2021 and 1 September 2021. For Singapore, the training period is between 1 April 2021 and 1 December 2021. For Malaysia, the training period is between 1 May 2021 and 1 October 2021.

\begin{figure}[H]
    \centering 
\makebox[\textwidth]{
\includegraphics[width=18cm]{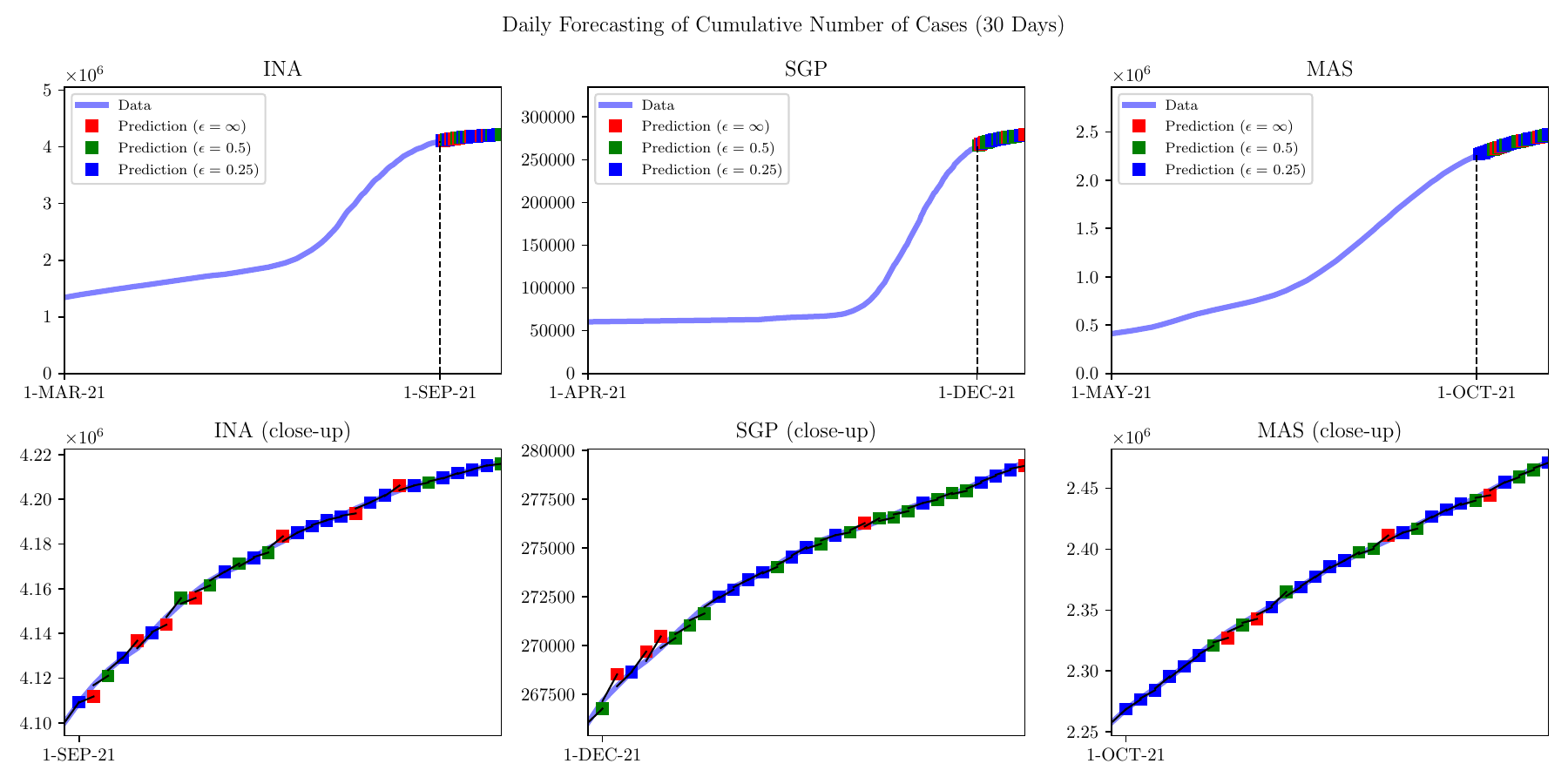}}
    \caption{The red squares indicate predictions that do not satisty the bound (\ref{bound}) under $\epsilon=0.5$. The green squares are predictions that satisfy the bound under $\epsilon=0.5$, but not under $\epsilon=0.25$. The blue squares are the predictions that satisfy the bound under both $\epsilon=0.25, \: 0.5$. The short black lines between any two consecutive days show the direction from the real data in a particular day to the prediction value in the next day.}
\label{fig:total_case_FORECAST}
\end{figure}

Given the data of the state variables $S(t)$, $I(t)$, and the cumulative number of cases $C(t)$, for every $t \in \{ t_{end}, \hdots,  t_{end}+29 \}$, we predict the total number of cases the next day $\widehat{C}(t+\Delta t)$, which are computed using the Runge-Kutta 4 numerical method. We perform the Runge-Kutta 4 numerical method at each daily interval $[t, t+\Delta t]$, for all $t \in \{ t_{end}, \hdots,  t_{end}+29 \}$, with real data $S(t)$, $I(t)$, and $C(t)$ as the initial conditions. The predictions are plotted in Figure \ref{fig:total_case_FORECAST}. Between two consecutive days, the absolute change in the real data is $|C(t+\Delta t) - C(t)|$. We categorize the predictions based on the following bound: 
\begin{equation}\label{bound}
|\widehat{C}(t+\Delta t)-C(t+\Delta t)| \le \epsilon |C(t+\Delta t) - C(t)|, \:\: \epsilon > 0.
\end{equation}
The smaller $\epsilon$, the more accurate the prediction that satisfies the bound (which depends on $\epsilon$). We found that there are 23, 25, and 26 predictions that fall inside the bound under $\epsilon=0.5$ for Indonesia, Singapore, and Malaysia, respectively. Moreover, there are 16, 12, and 17 predictions that fall in the bound under $\epsilon=0.25$ for Indonesia, Singapore, and Malaysia, respectively.

\section{Conclusion} \label{sec:conclusion}
\noindent We have introduced EPI-KAN to estimate time-dependent COVID-19 parameters based on historical data, Physic-Informed Neural Network, and the novel Kolmogorov-Arnold Network. We trained three different KAN-F functions: one for estimating transmission rates ($\widehat{\beta}$), one for estimating recovery rates ($\widehat{\gamma}$), and one for estimating mortality rates ($\widehat{\mu}$). We test two different network architectures: Model A that has 8 input features and Model B that has 4 input features. Under Model A, each network relies on 8 input variables, which consists of 4 state variables of $S^{rel}$, $I^{rel}$, $R^{rel}$, $D^{rel}$, and their 4 numerical gradients, all of which are normalized. While under Model B, the 4 numerical gradients are excluded as inputs. Our network models are able to estimate the time-dependent rates with decent accuracy. The accuracy of the estimation is measured by the accuracy of the data-fitting between the Runge-Kutta's numerical solution of the SIRD model (\ref{sird_model_S}-\ref{sird_model_D}) and the actual data. However, in the Singapore case, we discover that Model B has some difficulty in attaining appropriate model-fitting. One explanation may be the volatility of the number of active cases in Singapore. The dynamics of COVID-19 active cases in Singapore is patently not as smooth as in the other two countries. The study in \cite{fourier_mlp_google} also suggests that a network with low dimensional domain (small number of input variables) is known to have obstacles in fitting high frequency functions or data.

In addition, using KAN-F Model B, we can use information (real data) over the forecast period to predict the transmission rate that will occur between today and the next day. However, for all three countries, we found that the original output of the trained KAN-F Model B in the forecast period resulted in an overestimation of the total number of cases. Original predictions are also not proportional to the estimated transmission rates during the training period. Hence, we adjusted the prediction output by a scaling factor so that the forecast output is proportional to the training output. In fact, this scaling also strongly improves the accuracy of the prediction.

This research can be improved in several ways. Although we have presented specific settings for the KAN-F that produce decently accurate network models, further studies can analyze the performance for different numbers of hidden layers and neurons, and also for different values of $F$. Instead of using the SIRD model, further studies may also modify and improve the compartmental differential equations to be more nuanced and estimate more number of time-dependent parameters simultaneously. Our forecast only shows the predictions during the period when the SARS-CoV-2 Delta variant (B.1.617.2) was predominant. Further research may attempt to predict the number of cases during the period of the SARS-CoV-2 Omicron variant (B.1.1.529). Moreover, on can attempt to predict the dynamics of future active cases. This problem can be more intricate and nuanced, since it involves not only transmission rates, but also recovery and mortality rates, in the context of our SIRD model.

\section*{Declarations}
\subsection{Competing Interests}
\noindent The author has no known competing interests to declare.
\subsection{Code Availability}
\noindent We use Pytorch \cite{pytorch_site} to minimize the objective loss function (\ref{basic_loss_function}) and to obtain the estimated rates. The code and data associated with the result in this paper are available on GitHub at \url{https://github.com/anbarief/KAN_FOURIER_COVID19_MODELLING}.

\bibliographystyle{elsarticle-num} 
\bibliography{references}

\end{document}